\documentclass[prd,twocolumn,nofootinbib,superscriptaddress,amsmath,amssymb,floatfix]{revtex4-2}

\usepackage{graphicx}
\usepackage{booktabs}
\usepackage{bm}
\usepackage[colorlinks=true,citecolor=blue,linkcolor=red,urlcolor=blue]{hyperref}

\begin{document}

	\title{$X(2370)$ without exceptionalism from reanalyzing
$J/\psi \to \gamma + 3P$ data}
	
	\author{Tian-Cai Peng}
	\email{pengtc20@lzu.edu.cn}
	\affiliation{School of Physical Science and Technology, Lanzhou University, Lanzhou 730000, China}
	\affiliation{Lanzhou Center for Theoretical Physics, Key Laboratory of Theoretical Physics of Gansu Province, Key Laboratory of Quantum Theory and Applications of MoE, Gansu Provincial Research Center for Basic Disciplines of Quantum Physics, Lanzhou University, Lanzhou 730000, China}
	\affiliation{MoE Frontiers Science Center for Rare Isotopes, Lanzhou University, Lanzhou 730000, China}
	\affiliation{Research Center for Hadron and CSR Physics, Lanzhou University $\&$ Institute of Modern Physics of CAS, Lanzhou 730000, China}
	
	
	\author{Xiang~Liu}
	\email{xiangliu@lzu.edu.cn}
	\affiliation{School of Physical Science and Technology, Lanzhou University, Lanzhou 730000, China}
	\affiliation{Lanzhou Center for Theoretical Physics, Key Laboratory of Theoretical Physics of Gansu Province, Key Laboratory of Quantum Theory and Applications of MoE, Gansu Provincial Research Center for Basic Disciplines of Quantum Physics, Lanzhou University, Lanzhou 730000, China}
	\affiliation{MoE Frontiers Science Center for Rare Isotopes, Lanzhou University, Lanzhou 730000, China}
	\affiliation{Research Center for Hadron and CSR Physics, Lanzhou University $\&$ Institute of Modern Physics of CAS, Lanzhou 730000, China}
	
\begin{abstract}
Motivated by recent BESIII claims that $X(2370)$ is a pseudoscalar glueball candidate, we reanalyze several radiative $J/\psi$ decays into three pseudoscalars, $J/\psi\to\gamma PPP$, with emphasis on the $X(2370)$ region. A simultaneous analysis of the three invariant-mass projections in $J/\psi\to\gamma K_S^0K_S^0\eta^\prime$, including the pseudoscalar states $X(1835)$, $X(2100)$, $\eta(2225)$, $X(2370)$, $X(2500)$, and $\eta_c$, reproduces the observed dip near $2.2~\mathrm{GeV}$ and the enhancement near $2.4~\mathrm{GeV}$ through coherent interference, without invoking the additional broad $X(2800)$ component employed in the BESIII partial-wave analysis. We further analyze $J/\psi\to\gamma\pi^0\pi^0\eta$ and $J/\psi\to\gamma K_S^0K_S^0\pi^0$ by including both pseudoscalar and axial-vector contributions, and show that multiple excited meson states collectively describe the data. The absence of a prominent $X(2370)$ signal in the $K^*(892)$ region can be understood from differing orbital angular-momentum requirements rather than flavor-singlet suppression alone. Our results indicate that $X(2370)$ is not exceptional compared with the other reported pseudoscalar states in this mass region. Therefore, interpreting $X(2370)$ as a glueball-dominated state should be treated with caution.
\end{abstract}

	\maketitle

	\section{Introduction}
	
	The interpretation of $X(2370)$ as a glueball has recently attracted considerable attention and is the central focus of this work. This state was first reported in the $\eta^\prime\pi^+\pi^-$ invariant-mass spectrum of $J/\psi\to\gamma\eta^\prime\pi^+\pi^-$~\cite{BESIII:2010gmv}, in which several pseudoscalar-state candidates, including $X(1835)$, $X(2100)$, and $X(2600)$, have also been investigated.
	
	In 2005, a measurement of $J/\psi\to\gamma\eta^\prime\pi^+\pi^-$ based on a sample of 58 million $J/\psi$ events first revealed a new enhancement, $X(1835)$, without establishing any additional new states~\cite{BES:2005ega}. In 2011, with the accumulation of more data, an analysis of the same process based on 225 million $J/\psi$ events revealed two additional structures, $X(2120)$ and $X(2370)$, in the $\eta^\prime\pi^+\pi^-$ invariant-mass spectrum~\cite{BESIII:2010gmv}. Subsequently, the $X(2600)$ state was reported using a data sample of approximately 10 billion $J/\psi$ events~\cite{BESIII:2016fbr,BESIII:2022sfx}. This progression illustrates the importance of improved experimental precision and larger data samples in discovering new structures. Based on the observed states $X(1835)$, $X(2100)$, $X(2370)$, and $X(2600)$, explicit Regge trajectories can be constructed for the $\eta/\eta^\prime$ mesonic system, suggesting that $X(1835)$, $X(2100)$, and $X(2370)$ are plausible candidates for excited pseudoscalar mesons~\cite{Wang:2017iai}.
	
	Before 2024, the interpretation of $X(2370)$ as a glueball had not been widely pursued. One important reason was the considerable spread in lattice-QCD predictions for the mass of the pseudoscalar glueball, which ranged approximately from $2.3$ to $3.0~\mathrm{GeV}$~\cite{Bali:1993fb,Morningstar:1999rf,Chen:2005mg,Gregory:2012hu,Gui:2019dtm,Vadacchino:2023vnc}. In addition, the spin-parity quantum numbers of $X(2370)$ had not yet been determined.
	
	With the accumulation of experimental data, the BESIII Collaboration analyzed $J/\psi\to\gamma K_S^0K_S^0\eta^\prime$ using approximately 10 billion $J/\psi$ events in 2024, measuring the mass of $X(2370)$ to be $2.395\pm0.011\,{}^{+0.026}_{-0.094}~\mathrm{GeV}$ and determining its quantum numbers to be $J^{PC}=0^{-+}$~\cite{BESIII:2023wfi}. This mass is consistent with the value of $2.395(14)~\mathrm{GeV}$ predicted for the $0^{-+}$ glueball, $G_{0^{-+}}$, in a lattice-QCD calculation performed in 2019 under the quenched approximation~\cite{Gui:2019dtm}. Reference~\cite{Gui:2019dtm} also predicted a branching fraction of $(2.31\pm0.80)\times10^{-4}$ for $J/\psi\to\gamma G_{0^{-+}}$, which suggests that a pseudoscalar glueball may have a relatively large production rate in the $J/\psi$ radiative decay process. For this reason, $X(2370)$ has increasingly been interpreted as a pseudoscalar glueball candidate.
	
	Pseudoscalar glueballs are expected to exhibit gluon-mediated decay patterns that may resemble those of the pseudoscalar charmonium $\eta_c$~\cite{Chao:1995hd,Huang:1995td}. In particular, both are expected to decay into the same three-pseudoscalar-meson ($PPP$) final states. Based on this conjecture, BESIII continued to search for the $X(2370)$ signal in two other $J/\psi$ radiative decays, $J/\psi\to\gamma\pi^0\pi^0\eta$ and $J/\psi\to\gamma K_S^0K_S^0\pi^0$, using approximately 10 billion $J/\psi$ events~\cite{BESIII:2026rzt}. BESIII thus claimed that $X(2370)$ and $\eta_c$ are present in four favorable $PPP$ decay modes: $\eta^\prime\pi^+\pi^-$, $K_S^0K_S^0\eta^\prime$, $K_S^0K_S^0\pi^0$, and $\pi^0\pi^0\eta$. In particular, the BESIII analysis emphasized that in the $K_S^0K_S^0\eta^\prime$, $K_S^0K_S^0\pi^0$, and $\pi^0\pi^0\eta$ final states, only $X(2370)$ is observed together with $\eta_c$, which was argued to support the exceptionality of $X(2370)$ relative to the previously reported pseudoscalar states $X(1835)$, $X(2100)$, and $X(2600)$.
	
	In 2026, BESIII published its latest measurement of the $K^*(892)^0\bar K^0$ invariant-mass spectrum in $J/\psi\to\gamma K^*(892)^0\bar K^0$~\cite{BESIII:2026mvn}. Their interpretation is based on the argument that a pseudoscalar glueball is a typical flavor singlet, which is forbidden to couple to $K^*(892)\bar K$ by generalized G-parity conservation~\cite{Lipkin:1981ak}. Since no $X(2370)$ signal is observed in the $K^*(892)^0\bar K^0$ invariant-mass spectrum in $J/\psi\to\gamma K^*(892)^0\bar K^0$~\cite{BESIII:2026mvn}, BESIII claimed that $X(2370)$ is a flavor singlet.
	
	To date, BESIII has sought to establish a complete chain of logic through its series of studies, in order to regard $X(2370)$ as a pseudoscalar glueball-dominated state. We believe that the glueball picture of $X(2370)$ advocated by BESIII, and the arguments supporting it, are now clear to the reader~\cite{BESIII:2026rzt,BESIII:2026mvn}.

 We notice several recent theoretical studies, which have explored alternative interpretations of $X(2370)$, including hybrid, tetraquark, trigluon glueball, conventional $q\bar{q}$, and mixed configurations~\cite{Zhang:2026poz,Lu:2026foy}, $\bar{\Sigma}\Sigma$ molecular states~\cite{Ben:2026afy}, $\phi h_1(1380)$ dynamics~\cite{Khemchandani:2026syv}, and coupled-channel dynamics~\cite{Dong:2026smv}, highlighting the ambiguity in identifying its glueball dominant explanation.

In this work, we indicate that the experimental data allow for an understanding different from that in Refs.~\cite{BESIII:2026rzt,BESIII:2026mvn}, one supported by hadron spectroscopy.
	In this work, we reanalyze the $J/\psi\to\gamma K_S^0K_S^0\eta^\prime$ process based on our knowledge of pseudoscalar-meson spectroscopy~\cite{Wang:2017iai}. We find that, by including the contributions of $X(1835)$, $X(2100)$, $\eta(2225)$, $X(2370)$, $X(2500)$, and $\eta_c$, the $K_S^0K_S^0\eta^\prime$ line shape and the corresponding data can be well described. In particular, the pronounced dip around $2.2~\mathrm{GeV}$ and the enhancement around $2.4~\mathrm{GeV}$ are reproduced without introducing the additional effective broad $X(2800)$ component employed in the BESIII partial-wave analysis~\cite{BESIII:2023wfi}.
	
	Along this line, we reanalyze the $J/\psi\to\gamma\pi^0\pi^0\eta$ and $J/\psi\to\gamma K_S^0K_S^0\pi^0$ processes by introducing the allowed states in the pseudoscalar and axial-vector meson families. The present study differs substantially from the BESIII analyses in Refs.~\cite{BESIII:2026rzt,BESIII:2026mvn}: those analyses focus on local fits, whereas our study performs a full-range fit over the energy region covered by the corresponding invariant-mass spectra.
	
	Our analysis therefore shows that $X(2370)$ is not exceptional among the pseudoscalar states in this mass region, contrary to the BESIII claim~\cite{BESIII:2026rzt,BESIII:2026mvn}.
	
	In addition, we discuss the absence of an $X(2370)$ signal in the $K^*(892)$ signal region of $J/\psi\to\gamma K_S^0K_S^0\pi^0$, which should not be taken as strong evidence that $X(2370)$ is a flavor singlet. A detailed discussion of this issue is given in the following section.

	In general, we hope that our efforts will encourage caution in interpreting $X(2370)$ as a glueball. This is a practical application of Occam's razor: entities should not be multiplied beyond necessity.

	This paper is organized as follows. In Sec.~\ref{spectroscopy}, we review the pseudoscalar $\eta/\eta^\prime$ and axial-vector $f_1$ meson spectra, which provide the spectroscopic basis for our analysis.
	Then we present a simultaneous analysis of three invariant-mass spectra in the $J/\psi\to\gamma K_S^0K_S^0\eta^\prime$ process in Sec.~\ref{III}, and investigate other decay modes such as $J/\psi\to\gamma\pi^0\pi^0\eta$ and $J/\psi\to\gamma K_S^0K_S^0\pi^0$ in Sec.~\ref{modes}, to claim that $X(2370)$ is not particularly special compared with other pseudoscalar states. 
	Finally, a summary is given in Sec.~\ref{summary}.

	\section{The pseudoscalar and axial-vector mesons spectroscopy}\label{spectroscopy}
	
In this work, we investigate several $J/\psi \to \gamma PPP$ decay processes related to the observation of $X(2370)$. These processes are described in terms of an intermediate resonance $X$, which is produced through the radiative decay $J/\psi \to \gamma X$ and subsequently decays into three pseudoscalar mesons via the strong interaction, $X \to PPP$.

Depending on the partial waves involved in the radiative production and the subsequent strong decay, different quantum-number assignments are possible for the intermediate state. In particular, we consider the $J^{PC}=0^{-+}$ and $1^{++}$ assignments. We focus on several $J/\psi \to \gamma PPP$ processes corresponding to the $X(2370)$ observation claimed by BESIII.

\begin{figure}[htbp]
		\begin{center}
			\includegraphics[width=1\columnwidth]{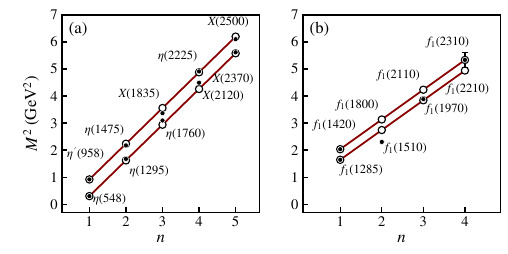}
			
			\caption{(a) Regge trajectories for the $\eta/\eta^\prime$ meson family ~\cite{Wang:2017iai}; (b) Regge trajectories for the $f_1$ meson family ~\cite{Chen:2015iqa}.}
			\label{fig:regge}
		\end{center}
	\end{figure}

	From a spectroscopic perspective, in Ref.~\cite{Wang:2017iai}, Wang \textit{et al.} constructed the $\eta/\eta^\prime$ family based on the experimentally observed pseudoscalar states and investigated their strong decay properties. Within this framework, $\eta(1295)$ and $\eta(1475)$ were assigned as the first radial excitations of $\eta$ and $\eta^\prime$, respectively, while $\eta(1760)$ and $X(1835)$ were considered candidates for the third radial excitations of the corresponding states. For higher excitations, $X(2100)$ and $\eta(2225)$ were proposed as candidates for $\eta(4S)$ and $\eta^\prime(4S)$, respectively, whereas $X(2370)$ and $X(2500)$ were suggested to correspond to $\eta(5S)$ and $\eta^\prime(5S)$. The Regge trajectories of the $\eta^\prime$ meson family are shown in Fig.~\ref{fig:regge}. These assignments were further examined through calculations of their strong decay widths, providing a systematic spectroscopic picture of the highly excited pseudoscalar meson family~\cite{Wang:2017iai}.

Later, motivated by the event accumulation around 2.6~GeV in the $\eta^\prime\pi^+\pi^-$ invariant-mass spectrum of $J/\psi\to\gamma\eta^\prime\pi^+\pi^-$ reported by the BESIII Collaboration~\cite{BESIII:2016fbr}, further studies extended the investigation to even higher radial excitations, providing useful theoretical guidance for future experimental searches~\cite{Wang:2020due}. These theoretical studies provide a systematic reference for identifying possible pseudoscalar resonances in the highly excited mass region.

	Compared with the pseudoscalar meson family, experimental information on axial-vector mesons with $J^{PC}=1^{++}$ is relatively limited. In Ref.~\cite{Chen:2015iqa}, Chen \textit{et al.} systematically investigated the mass spectrum and strong decay properties of the $f_1$ meson family, whose Regge trajectories are shown in Fig.~\ref{fig:regge}.

\begin{table}[htbp]
		\caption{Masses and widths of the pseudoscalar mesons and $\eta_c$.}
		\label{tab:resonance_input}
		\begin{ruledtabular}
			\begin{tabular}{lcc}
				State & Mass (MeV) & Width (MeV) \\
				\colrule
				$\eta$ & $547.862 \pm {0.017}$~\cite{ParticleDataGroup:2024cfk} & $0.00131  \pm 0.00005$~\cite{ParticleDataGroup:2024cfk} \\
				$\eta^\prime$ & $957.78\pm{0.06}$~\cite{ParticleDataGroup:2024cfk} & $0.1888\pm{0.006}$~\cite{ParticleDataGroup:2024cfk} \\
				
				$\eta(1295)$ & $1294 \pm 4$~\cite{ParticleDataGroup:2024cfk} & $55 \pm 5$~\cite{ParticleDataGroup:2024cfk} \\
				
				$\eta(1475)$ & $1476 \pm 4$~\cite{ParticleDataGroup:2024cfk} & $96 \pm 9$~\cite{ParticleDataGroup:2024cfk} \\
				$\eta(1760)$ & $1751 \pm 15$~\cite{ParticleDataGroup:2024cfk} & $240 \pm 30$~\cite{ParticleDataGroup:2024cfk} \\
				
				$X(1835)$
				& $1844\pm9{}^{+16}_{-25}$~\cite{BESIII:2015xco}
				& $192^{+20+62}_{-17-43}$~\cite{BESIII:2015xco} \\
				
				$X(2100)$
				& $2050^{+30+75}_{-24-26}$~\cite{BESIII:2016qzq}
				& $250^{+36+181}_{-30-164}$~\cite{BESIII:2016qzq} \\
				
				$\eta(2225)$
				& $2221^{+13}_{-10}$~\cite{ParticleDataGroup:2024cfk}
				& $185^{+40}_{-20}$~\cite{ParticleDataGroup:2024cfk} \\
				
				$X(2370)$
				& $2377\pm9$~\cite{ParticleDataGroup:2024cfk}
				& $148^{+80}_{-28}$~\cite{ParticleDataGroup:2024cfk} \\
				
				$X(2500)$
				& $2470^{+15+101}_{-19-23}$~\cite{BESIII:2016qzq}
				& $230^{+64+56}_{-35-33}$~\cite{BESIII:2016qzq} \\
				
				
				\hline
				$\eta_c$ & $2984.09 \pm 0.30$~\cite{ParticleDataGroup:2024cfk}
				& $30.0 \pm 0.5$~\cite{ParticleDataGroup:2024cfk} \\
			\end{tabular}
		\end{ruledtabular}
	\end{table}

Based on decay analysis, $f_1(1420)$ is supported as a candidate for the $s\bar{s}$ partner of the ground state $f_1(1285)$, while $f_1(1510)$ was considered a possible candidate for its first radial excitation. Furthermore, $f_1(1800)$, $f_1(1970)$, $f_1(2110)$, $f_1(2210)$, and $f_1(2310)$ were proposed as candidates for higher radial excitations and their corresponding strange partners.

\begin{table}[htbp]
		\caption{Masses and widths of the axial-vector resonances considered in the analysis.}
		\label{tab:axial}
		\begin{ruledtabular}
			\begin{tabular}{lcc}
				State & Mass (MeV) & Width (MeV) \\
				\colrule
				$f_1(1285)$ & $1281.8 \pm 0.5$~\cite{ParticleDataGroup:2024cfk} & $23 \pm 1.0$~\cite{ParticleDataGroup:2024cfk} \\
				$f_1(1420)$ & $1427.7^{+1.7}_{-1.4}$~\cite{ParticleDataGroup:2024cfk} & $55.4 \pm 3.2$ ~\cite{ParticleDataGroup:2024cfk}\\
				$f_1(1510)$ & $1518 \pm 5$~\cite{ParticleDataGroup:2024cfk} & $73 \pm 25$~\cite{ParticleDataGroup:2024cfk} \\
				$f_1(1970)$ & $1971 \pm 15 $~\cite{Anisovich:2000ut} & $240 \pm 45$~\cite{Anisovich:2000ut} \\
				$f_1(2310)$ & $2310 \pm 60 $~\cite{Anisovich:2000ut} & $255 \pm 70$~\cite{Anisovich:2000ut} \\
			\end{tabular}
		\end{ruledtabular}
	\end{table}

Since some of the predicted states have not yet been observed experimentally, we include only those $f_1$ states for which mass and width information has been reported in the literature, and the resonance parameters are summarized in Table~\ref{tab:axial}.
Guided by these two spectroscopic frameworks, we include the corresponding pseudoscalar and axial-vector states in the relevant mass region, as well as $\eta_c$, in the following analysis.

	\section{Reanalysis of the $J/\psi\to\gamma K_S^0K_S^0\eta^\prime$ process supported by pseudoscalar meson spectroscopy}\label{III}
	
	In 2024, the BESIII Collaboration performed a partial-wave analysis of $J/\psi\to\gamma K_S^0K_S^0\eta^\prime$ using a sample of $(10087\pm44)\times10^6$ $J/\psi$ events~\cite{BESIII:2023wfi}, and the spin-parity quantum numbers of $X(2370)$ were determined to be $J^{PC}=0^{-+}$.
The BESIII data used in this analysis are restricted to $m_{K_S^0K_S^0}<1.1$~GeV~\cite{BESIII:2023wfi}. Within this region, the two-dimensional distribution of $m_{K_S^0K_S^0}$ versus $m_{K_S^0K_S^0\eta^\prime}$ exhibits a pronounced near-threshold enhancement associated with the $f_0(980)$. This motivates us to consider the quasi-two-body decay chain $J/\psi\to\gamma X$, followed by $X\to f_0(980)\eta^\prime$ and $f_0(980)\to K_S^0K_S^0$.

Since the BESIII measurement of the photon polar-angle distribution in the $J/\psi$ rest frame supports a $J^{PC}=0^{-+}$ assignment for the intermediate resonance $X$~\cite{BESIII:2023wfi}, we restrict the resonant contributions in the present analysis to pseudoscalar states in the relevant mass region.

	The $J/\psi\to\gamma K_S^0K_S^0\eta^\prime$ process provides three invariant-mass projections: $m_{K_S^0K_S^0\eta^\prime}$, $m_{K_S^0K_S^0}$, and $m_{K_S^0\eta^\prime}$. They offer complementary information about the underlying decay dynamics and are described by a common amplitude model. The bin widths and relative normalization factors used for the three projections are summarized in Table~\ref{tab:fit_setup}, with $r_1=1$ taken as the reference normalization.

	\begin{table}[htbp]
		\caption{Numerical setup for the three projected spectra.}
		\label{tab:fit_setup}
		\begin{ruledtabular}
			\begin{tabular}{lcc}
				Spectrum & Bin width (GeV)  & Relative scale \\
				\colrule
				$m_{K_S^0K_S^0\eta^\prime}$ & $\Delta_1=0.015$ & $r_1=1$(fixed) \\
				$m_{K_S^0K_S^0}$ & $\Delta_2=0.002$  & $r_2=\frac{\Delta_2}{\Delta_1}=0.13$ \\
				$m_{K_S^0\eta^\prime}$ & $\Delta_3=0.018$& $r_3=\frac{\Delta_3}{\Delta_1}=1.20$ \\
			\end{tabular}
		\end{ruledtabular}
	\end{table}

\begin{figure}[htbp]
		\centering
		\includegraphics[width=\columnwidth]{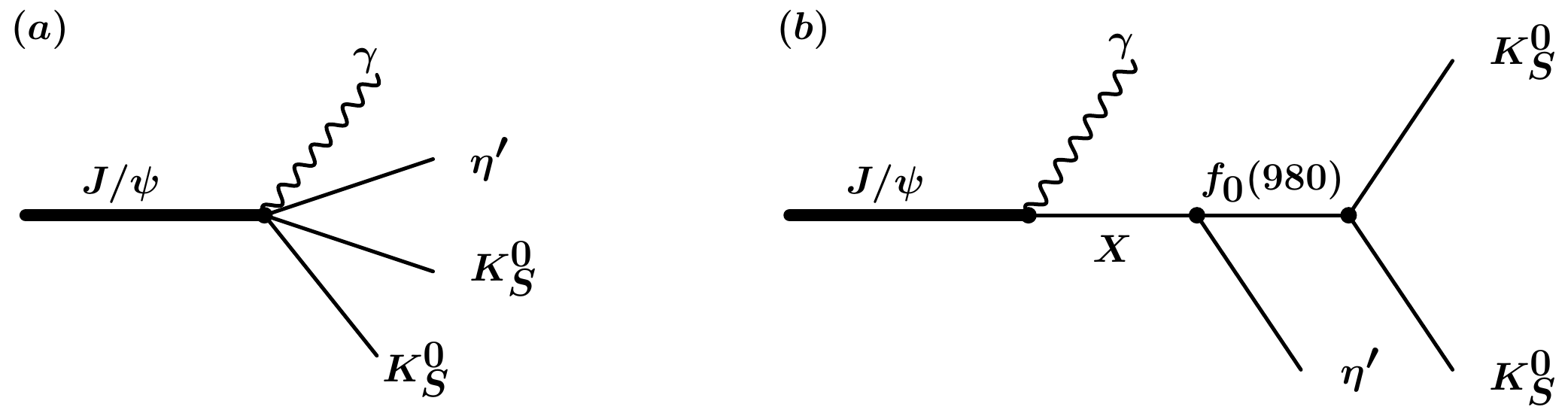}
		\caption{Schematic representations of the two contributions considered in $J/\psi\to\gamma K_S^0K_S^0\eta^\prime$: (a) nonresonant production of the final state and (b) resonant production through $J/\psi\to\gamma X$, followed by $X\to f_0(980)\eta^\prime$ and $f_0(980)\to K_S^0K_S^0$.}
		\label{fig:feynman}
	\end{figure}

	As illustrated in Fig.~\ref{fig:feynman}, two classes of contributions are considered. Panel (a) represents the nonresonant production mechanism, in which the $\gamma K_S^0K_S^0\eta^\prime$ final state is produced without an intermediate pseudoscalar resonance. Panel (b) illustrates the resonant mechanism proceeding through the intermediate states $X$ and $f_0(980)$. The two mechanisms lead to the same final state and are thus combined coherently at the amplitude level.
	
	The resonant amplitude is parameterized as a coherent sum of Breit--Wigner amplitudes, each multiplied by the $f_0(980)$ line shape:
	\begin{align}
		\mathcal A_{\rm res}
		&(m_{K_S^0K_S^0\eta^\prime},m_{K_S^0K_S^0})
		\nonumber\\
		=&\,F_{f_0}(m_{K_S^0K_S^0})
		\sum_i b_i e^{-i\phi_i}
		R_i(m_{K_S^0K_S^0\eta^\prime}),
		\label{eq:resamp}
	\end{align}
	with
	\begin{equation}\label{resonance}
		R_i(M)=
		\frac{1}{M^2-m_i^2+i m_i\Gamma_i},
		\quad
		M\equiv m_{K_S^0K_S^0\eta^\prime}.
	\end{equation}
	Here, $m_i$ and $\Gamma_i$ are the mass and width of the $i$-th pseudoscalar component, while $b_i$ and $\phi_i$ are its amplitude coefficient and relative phase, respectively. The $K_S^0K_S^0$ subsystem is described by the $f_0(980)$ Flatt\'e line shape~\cite{Flatte:1976xu},
	\begin{align}
		F_{f_0}(m_{K_S^0K_S^0})=1\Big/\Big[&m_{f_0}^2-m_{K_S^0K_S^0}^2 -i\bigl(g_1\rho_{\pi\pi}(m_{K_S^0K_S^0})\nonumber\\
		&+g_2\rho_{K_S^0K_S^0}(m_{K_S^0K_S^0})\bigr)\Big] ,
		\label{eq:flatte}
	\end{align}
	where
	\begin{align}
		\rho_{\pi\pi}(m_{K_S^0K_S^0})
		&=\sqrt{1-\frac{4m_\pi^2}{m_{K_S^0K_S^0}^2}},
		\nonumber\\
		\rho_{K_S^0K_S^0}(m_{K_S^0K_S^0})
		&=\sqrt{1-\frac{4m_{K_S^0}^2}{m_{K_S^0K_S^0}^2}}.
	\end{align}
	Here, $m_\pi$ denotes the pion mass, and the analytic continuation of the phase-space factors below the corresponding two-body thresholds is understood. The numerical parameters used for the Flatté line shape are $m_{f_0}=965\pm8\pm6$~MeV, $g_1=0.165\pm0.010\pm0.015$~GeV$^2$, and $g_2=4.21g_1$~\cite{BES:2004twe}.

The smooth nonresonant background amplitude corresponding to Fig.~\ref{fig:feynman}(a) is parametrized by a threshold-exponential form,
	\begin{equation}\label{bkg}
		\mathcal A_{\rm BKG}(x)=g x e^{-a\cdot x},\quad
		x=m_{K_S^0K_S^0\eta^\prime}-2m_{K_S^0}-m_{\eta^\prime},
	\end{equation}
	where $g$ and $a$ are phenomenological parameters. Since the resonant and nonresonant mechanisms lead to the same final state, their amplitudes are allowed to interfere. The total intensity is therefore given by
	\begin{align}
		I(m_{K_S^0K_S^0\eta^\prime},m_{K_S^0K_S^0})=\left|	\mathcal A_{\rm BKG} +\mathcal A_{\rm res} \right|^2.
		\label{eq:coherent}
	\end{align}

	Starting from the same amplitude, the three invariant-mass projections are presented. The theoretical bin contents for the three projections are then written as
	\begin{align}
    \frac{dN_1}{dm_{K_S^0K_S^0\eta^\prime}}
    &=r_1\Sigma_1(m_{K_S^0K_S^0\eta^\prime}),
    \nonumber\\
    \frac{dN_2}{dm_{K_S^0K_S^0}}
    &=r_2\Sigma_2(m_{K_S^0K_S^0}),
    \nonumber\\
    \frac{dN_3}{dm_{K_S^0\eta^\prime}}
    &=r_3\Sigma_3(m_{K_S^0\eta^\prime}).
    \end{align}
	where $r_1$, $r_2$, and $r_3$ are normalization factors for the corresponding projections. The three projected distributions are then used to constrain the common amplitude, and the details of the three invariant-mass spectrum projections ($\Sigma_{1,2,3}$) are presented in Appendix~\ref{app0}.


	With these ingredients, we perform a simultaneous fit to the three invariant-mass projections. In this energy range, the six pseudoscalar components considered are $X(1835)$, $X(2100)$, $\eta(2225)$, $X(2370)$, $X(2500)$, and $\eta_c$, whose masses and widths are listed in Table~\ref{tab:resonance_input}.

Since all three projections are obtained from the same event sample and are thus statistically correlated, the sum of the individual $\chi_i^2$ values is used only as a descriptive measure of the overall fit quality:
\begin{equation}
	\chi^2=
	\sum_{i=1}^{3}\sum_j
	\left[
	\frac{N_{ij}^{\rm exp}-N_{ij}^{\rm th}}
	{\delta N_{ij}^{\rm exp}}
	\right]^2,
	\label{eq:chi2}
\end{equation}
where $i=1,2,3$ labels the three invariant-mass projections and $j$ runs over the bins of the corresponding projection. The nonresonant background parameters $g$ and $a$, the amplitude coefficients $b_i$, and the relative phases $\phi_i$ are determined from the simultaneous fit. The resulting fit parameters are summarized in Table~\ref{tab:fit_parameters}. The contributions to $\chi^2$ from the $m_{K_S^0K_S^0\eta^\prime}$, $m_{K_S^0K_S^0}$, and $m_{K_S^0\eta^\prime}$ projections are 72.32, 77.80, and 43.15, respectively, giving a total $\chi^2=193.27$.

	\begin{table}[htbp]
		\begin{ruledtabular}
			\centering
			\caption{Fitted background parameters, amplitude coefficients, and relative phases.}
			\label{tab:fit_parameters}
			\begin{tabular}{cc|cc}
				{Parameter} & {Value} & {Parameter} & {Value} \\
				\midrule
				$g$ & $248.21 \pm 2.35$
				& $a$ & $1.10 \pm 0.01$ \\
				
				\hline
				
				$b_{X(1835)}$ & $86.78 \pm 0.46$
				& $\phi_{X(1835)}$ (rad) & $5.17 \pm 0.01$ \\
				
				$b_{X(2100)}$ & $23.21 \pm 0.60$
				& $\phi_{X(2100)}$ (rad) & $0.05 \pm 0.01$ \\
				
				$b_{\eta(2225)}$ & $6.72 \pm 0.18$
				& $\phi_{\eta(2225)}$ (rad) & $0.82 \pm 0.05$ \\
				
				$b_{X(2370)}$ & $14.18 \pm 0.15$
				& $\phi_{X(2370)}$ (rad) & $3.59 \pm 0.02$ \\
				
				$b_{X(2500)}$ & $31.13 \pm 0.22$
				& $\phi_{X(2500)}$ (rad) & $1.74 \pm 0.01$ \\
				
				$b_{\eta_c}$ & $5.76 \pm 0.15$
				& $\phi_{\eta_c}$ (rad) & $5.72 \pm 0.13$ \\
			\end{tabular}
		\end{ruledtabular}
	\end{table}
	
	\begin{figure*}[htbp]
		\centering
		\includegraphics[width=0.3\textwidth]{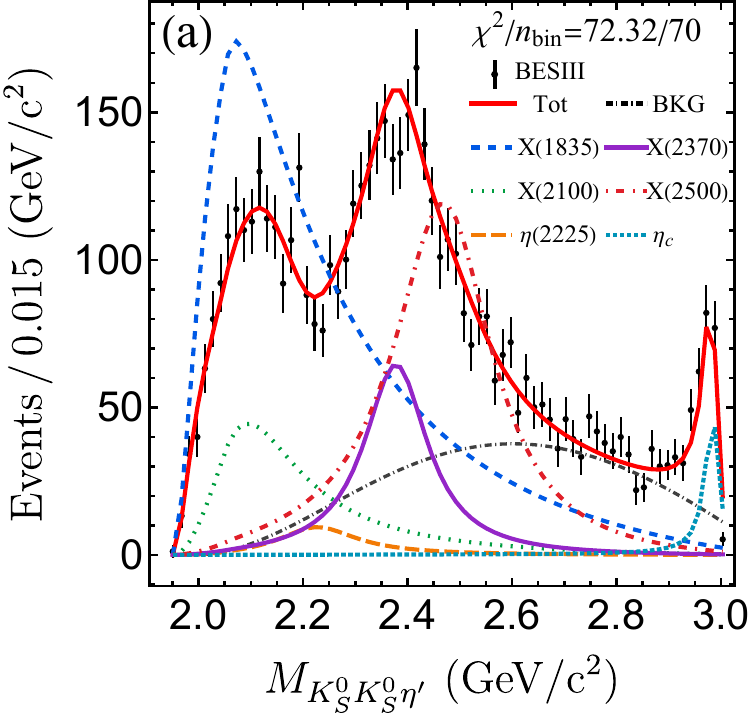}~~
		\includegraphics[width=0.31\textwidth]{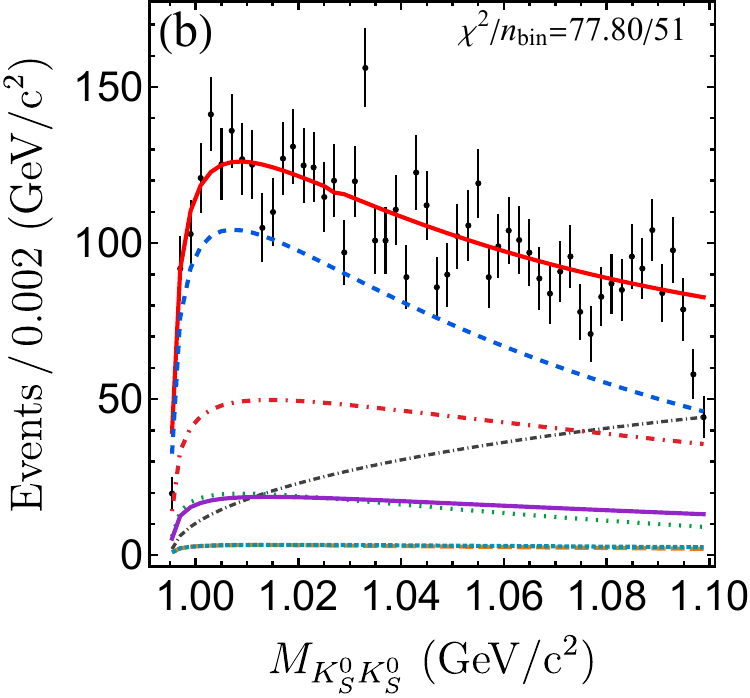}~~
		\includegraphics[width=0.3\textwidth]{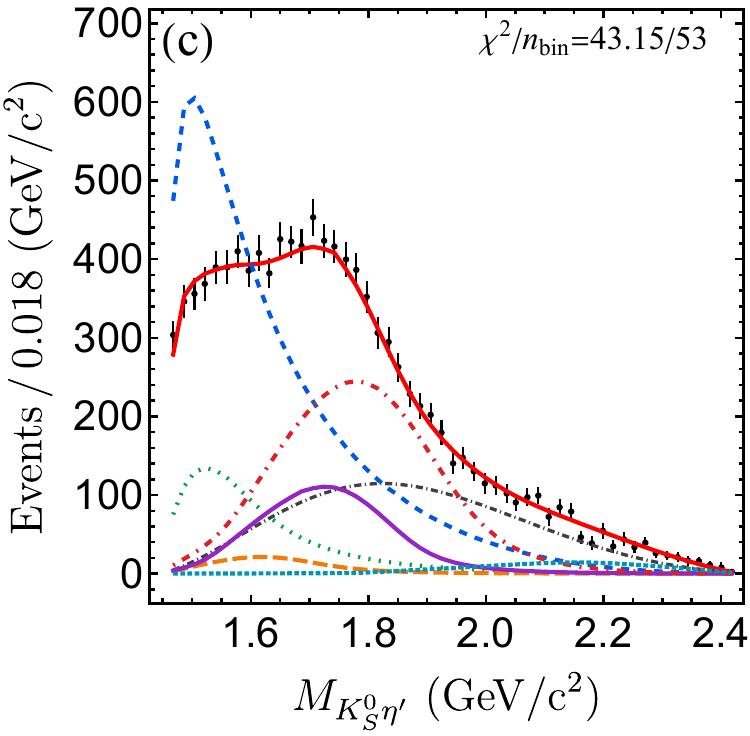}
		\caption{Invariant-mass distributions for the decay process $J/\psi\to\gamma K_S^0K_S^0\eta^\prime$: (a) $m_{K_S^0K_S^0\eta^\prime}$, (b) $m_{K_S^0K_S^0}$, and (c) $m_{K_S^0\eta^\prime}$. The black points with error bars represent the BESIII experimental data~\cite{BESIII:2023wfi}. The red solid curves show the total fit results, the black dashed curves represent the non-resonant background contributions, and the remaining curves indicate the individual pseudoscalar resonance contributions.}
		\label{fig:fit}
	\end{figure*}

As shown in Fig.~\ref{fig:fit}, all three projections are well described. Notably, this simultaneous description is achieved without introducing the additional broad effective $X(2800)$ component adopted in the BESIII analysis~\cite{BESIII:2023wfi}. Coherent interference among the pseudoscalar amplitudes is essential for reproducing the pronounced dip around $2.2$~GeV, showing that the observed line shape reflects not only the individual resonance contributions but also their relative phases. The successful description of the three projections within a framework motivated by the excited $\eta/\eta^\prime$ spectrum shows that the observed structures can be understood without attributing the enhancement at $2.4$~GeV to a single isolated resonance.

	
	
	Overall, our analysis shows that a unified description of the published invariant-mass projections in terms of these pseudoscalar meson states exists, which differs substantially from the BESIII analysis in Ref.~\cite{BESIII:2023wfi}. Our study also indicates that $X(2370)$ is not exceptional compared with other pseudoscalar states. Thus, the BESIII claim, emphasized in their study of $J/\psi\to \gamma K_S^0 K_S^0\eta^\prime$, that $X(2370)$ is exceptional is analysis-dependent and therefore open to question.

	\section{Other $PPP$ decay modes}\label{modes}
		
	Along this line, we further investigate the $PPP$ decay processes $J/\psi\to\gamma\pi^0\pi^0\eta$ and $J/\psi\to\gamma K_S^0K_S^0\pi^0$ by performing phenomenological fits to the corresponding invariant-mass spectra over the full investigated energy regions. In our analysis, contributions from the relevant excited pseudoscalar and axial-vector states are incorporated to provide a comprehensive description of the observed spectral structures. This full-range fitting strategy accounts for the combined effects of multiple resonances and their possible interference across the entire energy region, rather than attributing individual enhancements to isolated resonances.

This scheme differs from the analyses presented by the BESIII Collaboration in Refs.~\cite{BESIII:2026mvn,BESIII:2026rzt}, which focus on local fits in selected mass regions. Such local analyses may overlook contributions from other resonances outside the fitted intervals and can consequently lead to model-dependent interpretations of the observed structures.

The analysis incorporates two types of amplitudes corresponding to the quantum numbers $J^{PC}=0^{-+}$ and $1^{++}$, respectively:
\begin{align}
	\mathcal{A}^{0^{-+}} =&\ 
	\mathcal{A}_{\mathrm{bkg}}^{0^{-+}}
	+ \sum_i b_i R_i e^{i\phi_i},\\
	\mathcal{A}^{1^{++}} =&\ 
	\mathcal{A}_{\mathrm{bkg}}^{1^{++}}
	+ \sum_j b_j R_j e^{i\phi_j}.
\end{align}
Phenomenologically, the resulting invariant-mass distribution is expressed as
\begin{align}
	\frac{dN}{dM}
	= \Phi(M)\left(
	|\mathcal{A}^{0^{-+}}|^2
	+ |\mathcal{A}^{1^{++}}|^2
	\right).
\end{align}
Here, $\mathcal{A}_{\mathrm{bkg}}^{0^{-+}}$ and $\mathcal{A}_{\mathrm{bkg}}^{1^{++}}$ denote the nonresonant background amplitudes associated with the corresponding quantum numbers. For each decay channel, they are described by the same threshold-exponential functional form as in Eq.~\eqref{bkg}, with the threshold variable defined according to the corresponding three-body final state. The terms $R_i$ and $R_j$ represent the resonance amplitudes of the intermediate states with $J^{PC}=0^{-+}$ and $1^{++}$, respectively, while $b_i$ ($b_j$) and $\phi_i$ ($\phi_j$) denote their amplitude coefficients and relative phases. Here, $M$ is the invariant mass of the three-pseudoscalar ($PPP$) system, and $\Phi(M)$ represents the corresponding phase-space factor.

	\subsection{$J/\psi\to\gamma\pi^0\pi^0\eta$ in the full mass region and the $a_0(980)$ signal region}
		
	Recently, the BESIII Collaboration reported a measurement of $J/\psi\to\gamma\pi^0\pi^0\eta$~\cite{BESIII:2026rzt} to investigate the resonance parameters and decay properties of $X(2370)$. Since point-by-point uncertainties could not be reliably extracted from the published figures, we do not evaluate the statistical uncertainties of the fitted parameters in the present analysis.

\begin{figure}[htbp]
		\centering
		\includegraphics[width=0.4\textwidth]{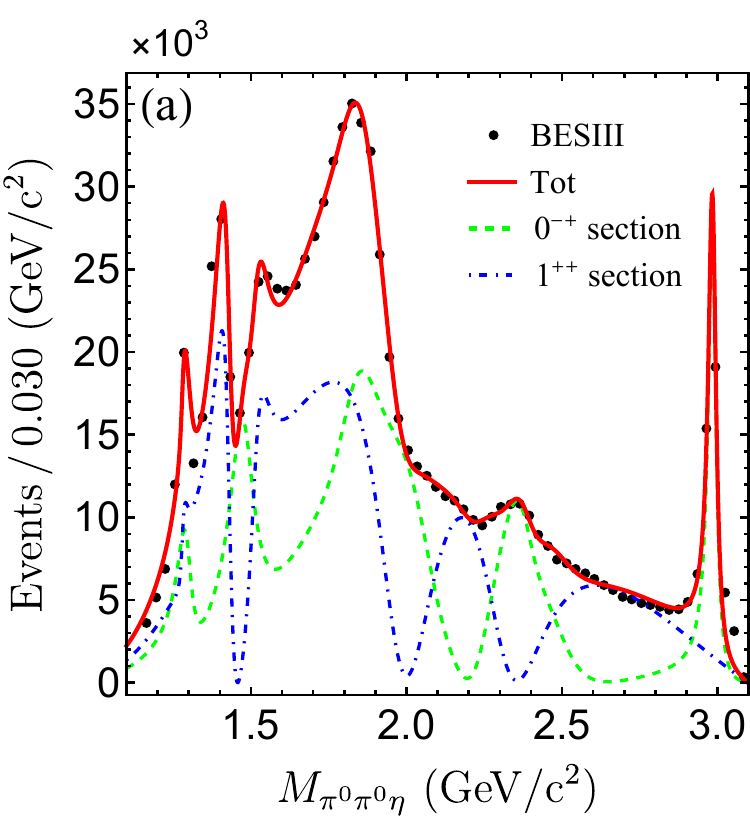} \\
		\includegraphics[width=0.4\textwidth]{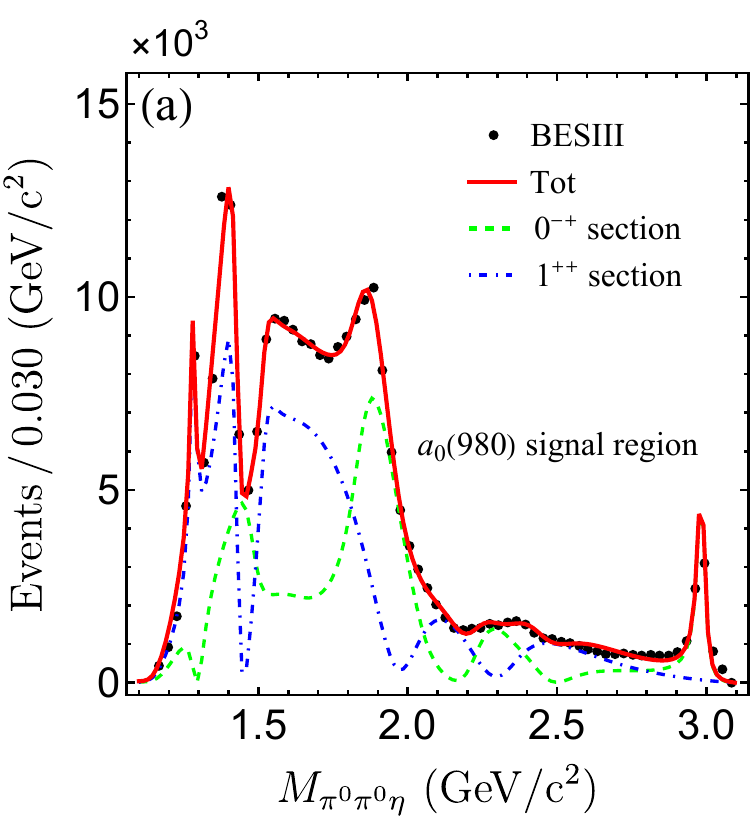}
		\caption{Invariant-mass distributions of the $\pi^0\pi^0\eta$ system in $J/\psi\to\gamma\pi^0\pi^0\eta$. (a) The full $\pi^0\pi^0\eta$ invariant-mass distribution. (b) The invariant-mass distribution in the $a_0(980)\to\pi^0\eta$ signal region, together with the fit results. }
		\label{fig:pipieta}
	\end{figure}

As shown in Fig.~\ref{fig:pipieta}, several distinct enhancement structures appear across the investigated energy region. This indicates that a comprehensive treatment of the relevant contributions is necessary to understand the observed spectral structures, rather than isolating $X(2370)$ and examining its decay properties independently.

Motivated by this observation, we extend our analysis by incorporating both $J^{PC}=0^{-+}$ and $1^{++}$ contributions into the $\pi^0\pi^0\eta$ invariant-mass spectrum, corresponding to the pseudoscalar and axial-vector states, respectively. The resulting fit is shown in Fig.~\ref{fig:pipieta}, and the fitted parameters are listed in Appendix~\ref{app}.
The combined contributions from these two components reproduce
the main features of both the full invariant-mass distribution
and the spectrum in the $a_0(980)$ signal region.
The corresponding fitted parameters and individual component
contributions are presented in Appendix~\ref{app}.
	
	The presence of multiple resonant states in this energy region indicates that constructive and destructive interference can substantially modify the observed spectral shapes and the apparent contributions of individual resonances, including $X(2370)$. Consequently, branching fractions extracted using a local fit with a simplified background parametrization may be sensitive to the assumed resonance content and fitting strategy and should be interpreted with caution rather than regarded as model-independent evidence for the decay properties of $X(2370)$.

More generally, contributions from neighboring excited meson states with different quantum numbers may play an important role in shaping the observed invariant-mass distributions but are not necessarily fully accounted for in analyses restricted to selected mass regions. In particular, the structures observed around 2.4~GeV may admit alternative interpretations when the contributions of additional excited states and their interference effects are taken into consideration. These observations underscore the importance of considering the excited-meson spectrum over the full energy region and adopting a full-range fitting strategy to obtain a more comprehensive understanding of the observed structures and the nature of $X(2370)$.

	\subsection{$J/\psi\to\gamma K_S^0K_S^0\pi^0$ process}

	In 2026, the BESIII Collaboration reported an analysis of $J/\psi\to\gamma K_S^0K_S^0\pi^0$~\cite{BESIII:2026mvn}. Motivated by this study, we perform a phenomenological analysis of the $K_S^0K_S^0\pi^0$ invariant-mass spectrum by incorporating contributions from both pseudoscalar ($J^{PC}=0^{-+}$) and axial-vector ($J^{PC}=1^{++}$) meson states over the full investigated energy region.
	
	\begin{figure}[htbp]
		\centering
		\includegraphics[width=0.4\textwidth]{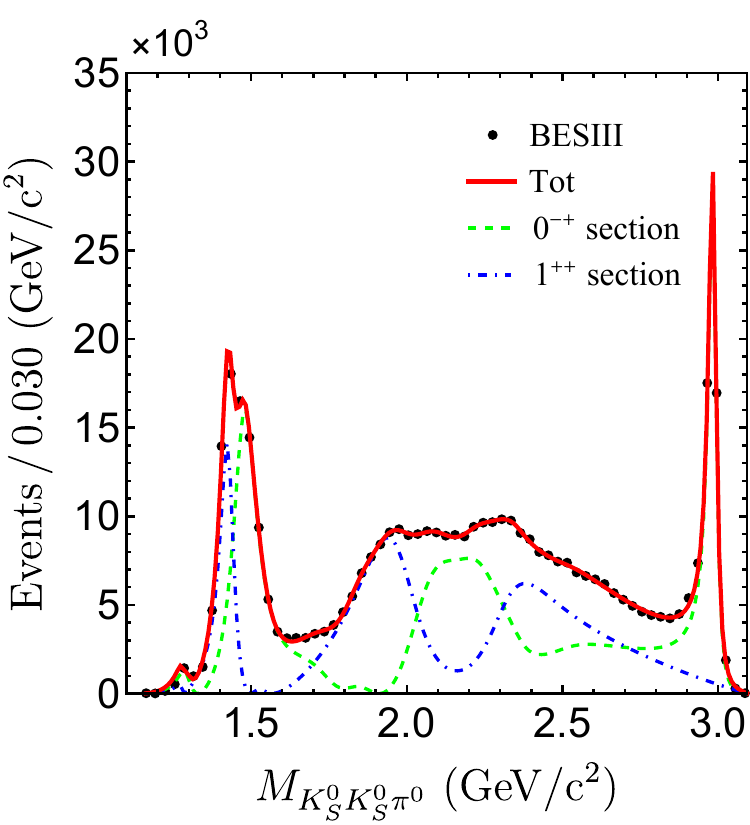}
		\caption{Invariant-mass distribution of the $K_S^0K_S^0\pi^0$ system in the $J/\psi\to\gamma K_S^0K_S^0\pi^0$ process, together with the corresponding fit result.}
		\label{fig:kkpi}
	\end{figure}
	
	As shown by the data points in Fig.~\ref{fig:kkpi}, the invariant-mass distribution in the energy region between 1.8 and 2.5~GeV exhibits a substantial deviation from the phase-space distribution, yet no clear and isolated resonance peak is observed.
    This behavior suggests that the observed structures may arise from the combined contributions of multiple intermediate states, involving broad resonances and their interference effects. Therefore, the enhancement around 2.4~GeV should not be attributed solely to a single $X(2370)$ resonance superimposed on a local polynomial background. Instead, contributions from other relevant excited meson states across the entire energy region should be taken into account.

As shown in Fig.~\ref{fig:kkpi}, both the $0^{-+}$ and $1^{++}$ components make sizable contributions to the observed invariant-mass spectrum. The individual contributions from these two components, together with the fitted parameters, are presented in Appendix~\ref{app}. By considering the full invariant-mass distributions, our analysis provides a more comprehensive description of the interplay among pseudoscalar and axial-vector states. Our results demonstrate that the observed spectral structures can be described by the combined contributions of pseudoscalar and axial-vector states, without treating $X(2370)$ as a standalone state.

	\subsection{$J/\psi\to\gamma K_S^0K_S^0\pi^0$ process in the $K^*(892)$ signal region }

Furthermore, the BESIII Collaboration performed a dedicated analysis of the $K^*(892)$ signal region in $J/\psi \to \gamma K_S^0 K_S^0 \pi^0$~\cite{BESIII:2026mvn}. As in their analyses of other $J/\psi\to\gamma PPP$ decay modes, a local fit was performed to the invariant-mass spectrum in the selected signal region. Notably, the absence of a prominent $X(2370)\to K^*(892)\bar{K}^0$ signal was interpreted as evidence for the flavor-singlet nature of $X(2370)$, thereby lending support to its possible glueball interpretation while disfavoring its conventional $q\bar{q}$ assignment.

However, the absence of a prominent $X(2370)$ signal in the $K^*(892)$ signal region does not necessarily imply flavor-singlet suppression. The sensitivity to the process $X\to K^*(892)\bar{K}$ depends on the angular-momentum configurations involved in both the production and decay processes. For a pseudoscalar state with $J^{PC}=0^{-+}$, both $J/\psi(1^{--})\to\gamma X(0^{-+})$ and $X(0^{-+})\to K^*(892)(1^{-})\bar{K}(0^{-})$ proceed through $P$-wave transitions, which may lead to dynamical suppression. In contrast, an axial-vector state with $J^{PC}=1^{++}$ can proceed through $S$-wave configurations in both processes, potentially making the corresponding channel more favorable for the observation of axial-vector contributions. This difference in angular-momentum requirements provides an alternative explanation for the absence of a prominent $X(2370)$ signal in this region.

\begin{figure}[htbp]
		\centering
		\includegraphics[width=0.4\textwidth]{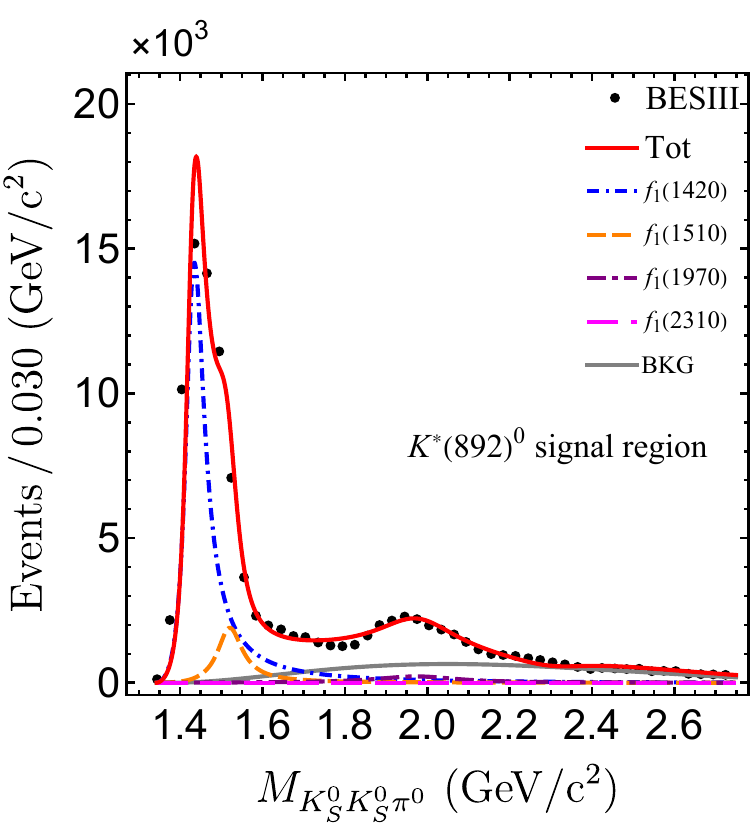}
		\caption{Invariant-mass distribution of $K_S^0K_S^0\pi^0$ in the $K^*(892)^0\to K_S^0\pi^0$ signal region of $J/\psi\to\gamma K_S^0K_S^0\pi^0$, together with the fit result.}
		\label{fig:kkpi_k892}
	\end{figure}

Motivated by this consideration, we investigate the $K_S^0K_S^0\pi^0$ invariant-mass distribution in the $K^*(892)^0$ signal region by considering contributions from axial-vector states, as shown in Fig.~\ref{fig:kkpi_k892}. The contributions from $f_1(1420)$, $f_1(1510)$, $f_1(1970)$, and $f_1(2310)$, together with a nonresonant background, provide a reasonable description of the observed structures around 1.4 and 2.0~GeV. In contrast, no significant contributions from pseudoscalar states, such as $X(1835)$, are needed to describe the spectrum in this region. Accordingly, we do not include the $0^{-+}$ component in the phenomenological description of the $K^*(892)$ signal region.

	Our analysis shows that the axial-vector ($J^{PC}=1^{++}$) contributions can account for the observed enhancement in the $K^*(892)$ signal region. This suggests that the absence of a prominent $X(2370)$ signal in this channel may be understood in terms of the different orbital angular-momentum requirements for pseudoscalar and axial-vector contributions, rather than being uniquely attributable to flavor-singlet suppression. Therefore, the observed behavior highlights the importance of considering partial-wave dynamics when interpreting the decay properties of $X(2370)$.

	\section{Summary}\label{summary}

Before 2024, the glueball interpretation of $X(2370)$ had attracted little attention, mainly because of the wide spread in lattice-QCD predictions for the pseudoscalar glueball mass and the lack of a determination of its spin-parity quantum numbers. In 2024, BESIII analyzed $J/\psi \to \gamma K_S^0 K_S^0 \eta^\prime$ with about 10 billion $J/\psi$ events and determined $J^{PC}=0^{-+}$ for $X(2370)$~\cite{BESIII:2023wfi}. The measured mass agrees with the 2019 lattice-QCD prediction of $2.395(14)$ GeV for the lightest pseudoscalar glueball $G_{0^{-+}}$~\cite{Gui:2019dtm}, renewing interest in the glueball interpretation.

Motivated by this interpretation, BESIII subsequently studied $X(2370)$ in other $PPP$ channels, including $J/\psi \to \gamma \pi^0\pi^0\eta$ and $J/\psi \to \gamma K_S^0 K_S^0\pi^0$~\cite{BESIII:2026rzt,BESIII:2026mvn}, and argued that its decay behavior distinguishes it from other pseudoscalar states in the same mass region. However, some of these analyses rely on local fits to restricted mass regions, which introduces considerable model dependence and limits a comprehensive understanding of the production and decay mechanisms of $X(2370)$.

In general, $J/\psi \to \gamma PPP$ processes may receive contributions from intermediate resonances with different quantum numbers, particularly $J^{PC}=0^{-+}$ and $1^{++}$. Motivated by previous studies of the pseudoscalar~\cite{Wang:2017iai} and axial-vector~\cite{Chen:2015iqa} meson spectra, we systematically investigate several relevant radiative $J/\psi$ decays by incorporating the corresponding pseudoscalar and axial-vector states.

First, we revisit $J/\psi \to \gamma K_S^0 K_S^0 \eta^\prime$ within pseudoscalar meson spectroscopy. Including $X(1835)$, $X(2100)$, $\eta(2225)$, $X(2370)$, $X(2500)$, and $\eta_c$, we obtain a satisfactory description of the $K_S^0 K_S^0 \eta^\prime$ invariant-mass spectrum and its projections. In particular, the dip around 2.2~GeV and the enhancement around 2.4~GeV are reproduced through coherent interference among these pseudoscalar contributions, without the additional broad effective $X(2800)$ component employed in the BESIII partial-wave analysis. The observed line shape is thus accommodated within the established and candidate excited pseudoscalar meson spectrum.

Along this line, we investigate $J/\psi \to \gamma \pi^0\pi^0\eta$ and $J/\psi \to \gamma K_S^0 K_S^0\pi^0$ by incorporating both pseudoscalar ($J^{PC}=0^{-+}$) and axial-vector ($J^{PC}=1^{++}$) contributions. Our analyses provide satisfactory descriptions of the corresponding invariant-mass distributions over the investigated energy regions. The results indicate that multiple excited pseudoscalar and axial-vector states play important roles in shaping the observed spectra; the enhancement structures therefore cannot necessarily be attributed to a single isolated resonance, and a comprehensive treatment of the relevant meson spectrum is essential.

Furthermore, we discuss the absence of a prominent $X(2370)$ signal in the $K^*(892)$ signal region of $J/\psi \to \gamma K_S^0 K_S^0\pi^0$. The observed invariant-mass distribution there can be reasonably described by axial-vector states, including $f_1(1420)$, $f_1(1510)$, $f_1(1970)$ and $f_1(2310)$, together with a nonresonant background. From angular-momentum conservation, the production and decay of a pseudoscalar state into $K^*(892)\bar{K}$ involve $P$-wave transitions, whereas the corresponding axial-vector processes can proceed through $S$-wave configurations. Such differences may lead to different dynamical sensitivities to the two types of intermediate states. Hence, the absence of a prominent $X(2370)$ signal in the $K^*(892)$ region should not be regarded as strong evidence for its flavor-singlet nature or as a definitive indication of its glueball interpretation.

In summary, the observed decay patterns of $X(2370)$ can be accommodated within a conventional meson spectroscopy framework involving multiple excited pseudoscalar and axial-vector states. The apparent exceptional behavior of $X(2370)$ may be influenced by the contributions and interference effects of other resonances in the relevant energy region. A comprehensive investigation of the excited meson spectrum across the full energy range is therefore essential before drawing definitive conclusions about the internal structure of $X(2370)$. In particular, caution should be exercised when interpreting $X(2370)$ as a glueball-dominated state, since the currently available experimental observations do not uniquely establish such an interpretation.
	
	\begin{acknowledgments}
		This work is supported by the Natural Science Foundation of Gansu Province (No. 26RCKA012 and No. 25JRRA799), the National Natural Science Foundation of China under Grants No. 12335001 and No. 12247101, the ``111 Center" under Grant No. B20063, the Fundamental Research Funds for the Central Universities (lzujbky-2023-stlt01), and Lanzhou City High-Level Talent Funding.
	\end{acknowledgments}

	\section*{Data availability}
 
	The experimental data points are extracted by digitizing the published figures.

	\bibliography{references.bib}

	\appendix

	\section{Three invariant-mass projections in $J/\psi\to\gamma K_S^0K_S^0\eta^\prime$}\label{app0}
	
	To construct the three invariant-mass projections from this common intensity $I(m_{K_S^0K_S^0\eta^\prime},m_{K_S^0K_S^0})$, the relevant kinematic factors and integration limits must be specified. The radiative-photon momentum in the $J/\psi$ rest frame is
	\begin{equation}
		p_\gamma(m_{K_S^0K_S^0\eta^\prime})=\frac{m_{J/\psi}^2-m_{K_S^0K_S^0\eta^\prime}^2}{2m_{J/\psi}}.
	\end{equation}
	For the $K_S^0K_S^0$ subsystem, the daughter-particle momenta are
	\begin{align}
		p_{\eta^\prime}(m_{K_S^0K_S^0\eta^\prime},m_{K_S^0K_S^0})&=\frac{\sqrt{\lambda(m_{K_S^0K_S^0\eta^\prime}^2,m_{K_S^0K_S^0}^2,m_{\eta^\prime}^2)}}{2m_{K_S^0K_S^0\eta^\prime}},\nonumber\\
		p_{K_S^0}(m_{K_S^0K_S^0})&=\frac{\sqrt{\lambda(m_{K_S^0K_S^0}^2,m_{K_S^0}^2,m_{K_S^0}^2)}}{2m_{K_S^0K_S^0}}.
	\end{align}
	where the K\"all\'en function is defined as
	\begin{equation}
		\lambda(x,y,z)=x^2+y^2+z^2-2xy-2xz-2yz.
	\end{equation}
	
	The first projection corresponds to the invariant mass $m_{K_S^0K_S^0\eta^\prime}$. Integrating over the allowed $K_S^0K_S^0$ mass range gives
	\begin{align}
		\Sigma_1(m_{K_S^0K_S^0\eta^\prime})=&\int_{2m_{K_S^0}}^{\min(m_{K_S^0K_S^0}^{\rm cut},\,m_{K_S^0K_S^0\eta^\prime}-m_{\eta^\prime})}\mathrm{d} m_{K_S^0K_S^0}\nonumber\\
		&\times p_\gamma(m_{K_S^0K_S^0\eta^\prime})\,p_{K_S^0}(m_{K_S^0K_S^0})\nonumber\\
		&\times p_{\eta^\prime}(m_{K_S^0K_S^0\eta^\prime},m_{K_S^0K_S^0})\nonumber\\
		&\times I(m_{K_S^0K_S^0\eta^\prime},m_{K_S^0K_S^0}),
		\label{eq:sigma1}
	\end{align}
	where $m_{K_S^0K_S^0}^{\rm cut}=1.1~\mathrm{GeV}$.  In this projection, one data point closest to the selection boundary is excluded. 
	
	The second projection corresponds to the invariant mass $m_{K_S^0K_S^0}$. In this case, the integration is performed over the allowed three-body invariant mass:
	\begin{align}
		\Sigma_2(m_{K_S^0K_S^0})=&\int_{m_{K_S^0K_S^0}+m_{\eta^\prime}}^{m_{J/\psi}}\mathrm{d} m_{K_S^0K_S^0\eta^\prime}\nonumber\\
		&\times p_\gamma(m_{K_S^0K_S^0\eta^\prime})\,p_{K_S^0}(m_{K_S^0K_S^0})\nonumber\\
		&\times p_{\eta^\prime}(m_{K_S^0K_S^0\eta^\prime},m_{K_S^0K_S^0})\nonumber\\
		&\times I(m_{K_S^0K_S^0\eta^\prime},m_{K_S^0K_S^0}).
		\label{eq:sigma2}
	\end{align}
	For this projection, the two points closest to the selection boundaries are excluded.
	
	The third projection corresponds to the invariant mass $m_{K_S^0\eta^\prime}$.
    For fixed values of $m_{K_S^0K_S^0\eta^\prime}$ and $m_{K_S^0\eta^\prime}$, the kinematically allowed range of $m_{K_S^0K_S^0}^2$ is
	\begin{equation}
		m_{K_S^0K_S^0,\min}^2
		\leq m_{K_S^0K_S^0}^2
		\leq m_{K_S^0K_S^0,\max}^2,
	\end{equation}
	where $m_{K_S^0K_S^0,\min/\max}^2 = 2m_{K_S^0}^2+2(E_1E_2\mp p_1p_2)$,
	and
	\begin{align}
		E_1&=
		\frac{m_{K_S^0\eta^\prime}^2
			+m_{K_S^0}^2-m_{\eta^\prime}^2}
		{2m_{K_S^0\eta^\prime}},
		\nonumber\\
		p_1&=
		\frac{
			\sqrt{\lambda(m_{K_S^0\eta^\prime}^2,
				m_{K_S^0}^2,m_{\eta^\prime}^2)}
		}{2m_{K_S^0\eta^\prime}},
		\nonumber\\
		E_2&=
		\frac{m_{K_S^0K_S^0\eta^\prime}^2
			-m_{K_S^0\eta^\prime}^2-m_{K_S^0}^2}
		{2m_{K_S^0\eta^\prime}},
		\nonumber\\
		p_2&=
		\frac{
			\sqrt{\lambda(m_{K_S^0K_S^0\eta^\prime}^2,
				m_{K_S^0\eta^\prime}^2,m_{K_S^0}^2)}
		}{2m_{K_S^0\eta^\prime}}.
	\end{align}
	
	Taking into account the selection requirement $m_{K_S^0K_S^0}<m_{K_S^0K_S^0}^{\rm cut}=1.1$~GeV, the integration limits are
	\begin{align}
		m_{K_S^0K_S^0,\rm low}
		&=\max\left[
		2m_{K_S^0},
		m_{K_S^0K_S^0,\min}
		\right],
		\nonumber\\
		m_{K_S^0K_S^0,\rm high}
		&=\min\left[
		m_{K_S^0K_S^0}^{\rm cut},
		m_{K_S^0K_S^0,\max}
		\right].
	\end{align}
	
	The projected distribution is then given by
	\begin{align}
		\Sigma_3(m_{K_S^0\eta^\prime})
		=&\,n_{K_S^0\eta^\prime}
		\int_{m_{K_S^0\eta^\prime}+m_{K_S^0}}^{m_{J/\psi}}
		\mathrm{d}m_{K_S^0K_S^0\eta^\prime}
		\nonumber\\
		&\times
		\int_{m_{K_S^0K_S^0,\rm low}}
		^{m_{K_S^0K_S^0,\rm high}}
		\mathrm{d}m_{K_S^0K_S^0}
		\nonumber\\
		&\times p_\gamma(m_{K_S^0K_S^0\eta^\prime})
		\frac{m_{K_S^0\eta^\prime}\cdot m_{K_S^0K_S^0}}
		{2m_{K_S^0K_S^0\eta^\prime}}
		\nonumber\\
		&\times
		I(m_{K_S^0K_S^0\eta^\prime},
		m_{K_S^0K_S^0}),
		\label{eq:sigma3}
	\end{align}
	where $n_{K_S^0\eta^\prime}=2$, since each event contributes two possible $K_S^0\eta^\prime$ combinations.

	\newpage
	
	~

	\newpage
	
	~

	\widetext{
		
		\section{Fitting parameters and individual component contributions for other $PPP$ decay modes}\label{app}
		
	}
	
	\begin{table*}[htbp]
		\centering
		\caption{Fitted background parameters, amplitude coefficients, and relative phases for the pseudoscalar and axial-vector components in the different decay modes. In the $K^*(892)$ signal region, the $0^{-+}$ component and $f_1(1285)$ are not included. The values listed here are the central values obtained from the phenomenological fits, the parameter uncertainties uncertainties are not quoted because reliable point-by-point experimental uncertainties are unavailable.}
		\label{tab:fit_4}
		\small
		\setlength{\tabcolsep}{5pt}
		\renewcommand{\arraystretch}{1.15}
		\begin{ruledtabular}
			\begin{tabular}{lcccc}
				Parameter
				& \multicolumn{2}{c}{$J/\psi\to\gamma\pi^0\pi^0\eta$}
				& \multicolumn{2}{c}{$J/\psi\to\gamma K_S^0K_S^0\pi^0$} \\
				\cline{2-3}\cline{4-5}
				& full region
				& $a_0(980)$ region
				& full region
				& $K^*(892)$ region \\
				\colrule
				
				\multicolumn{5}{c}{$J^{PC}=0^{-+}$~section} \\
				\hline

				$g_1$
				& $-24.75$
				& $-143.45$
				& $-20.84$
				& $-$ \\
				
				$a_1$
				& $1.52$
				& $2.74$
				& $1.55$
				& $-$ \\

				$b_{\eta(1295)}$
				& $0.59$
				& $0.95$
				& $1.31$
				& $-$ \\
				
				$b_{\eta(1475)}$
				& $1.18$
				& $0.80$
				& $3.38$
				& $-$ \\
				
				$b_{\eta(1760)}$
				& $7.13$
				& $1.27$
				& $2.70$
				& $-$ \\
				
				$b_{X(1835)}$
				& $3.31$
				& $4.42$
				& $2.14$
				& $-$ \\
				
				$b_{X(2100)}$
				& $2.78$
				& $3.25$
				& $3.30$
				& $-$ \\
				
				$b_{\eta(2225)}$
				& $1.59$
				& $3.76$
				& $0.79$
				& $-$ \\
				
				$b_{X(2370)}$
				& $1.41$
				& $0.39$
				& $2.34$
				& $-$ \\
				
				$b_{X(2500)}$
				& $2.45$
				& $4.21$
				& $0.67$
				& $-$ \\
				
				$b_{\eta_c}$
				& $1.38$
				& $2.83$
				& $1.40$
				& $-$ \\

				$\phi_{\eta(1295)}$ (rad)
				& $0.00$
				& $4.65$
				& $5.73$
				& $-$ \\
				
				$\phi_{\eta(1475)}$ (rad)
				& $0.00$
				& $0.06$
				& $0.58$
				& $-$ \\
				
				$\phi_{\eta(1760)}$ (rad)
				& $4.87$
				& $2.65$
				& $4.98$
				& $-$ \\
				
				$\phi_{X(1835)}$ (rad)
				& $2.19$
				& $3.03$
				& $0.49$
				& $-$ \\
				
				$\phi_{X(2100)}$ (rad)
				& $0.00$
				& $5.06$
				& $1.99$
				& $-$ \\
				
				$\phi_{\eta(2225)}$ (rad)
				& $4.16$
				& $4.13$
				& $1.12$
				& $-$ \\
				
				$\phi_{X(2370)}$ (rad)
				& $0.00$
				& $4.23$
				& $4.23$
				& $-$ \\
				
				$\phi_{X(2500)}$ (rad)
				& $6.28$
				& $5.72$
				& $4.29$
				& $-$ \\
				
				$\phi_{\eta_c}$ (rad)
				& $0.31$
				& $1.83$
				& $0.46$
				& $-$ \\
				
				\hline
				\multicolumn{5}{c}{$J^{PC}=1^{++}$~section} \\
				\hline

				$g_2$
				& $47.96$
				& $140.44$
				& $28.43$
				& $24.11$ \\
				
				$a_2$
				& $1.49$
				& $2.30$
				& $1.58$
				& $1.50$ \\

				$b_{f_1(1285)}$
				& $0.90$
				& $0.38$
				& $0.35$
				& $-$ \\
				
				$b_{f_1(1420)}$
				& $1.46$
				& $1.90$
				& $2.53$
				& $4.36$ \\
				
				$b_{f_1(1510)}$
				& $0.69$
				& $0.59$
				& $0.64$
				& $1.54$ \\
				
				$b_{f_1(1970)}$
				& $4.58$
				& $5.67$
				& $2.68$
				& $1.64$ \\
				
				$b_{f_1(2310)}$
				& $4.01$
				& $4.88$
				& $1.58$
				& $0.83$ \\

				$\phi_{f_1(1285)}$ (rad)
				& $1.13$
				& $4.16$
				& $1.84$
				& $-$ \\
				
				$\phi_{f_1(1420)}$ (rad)
				& $2.53$
				& $2.21$
				& $2.69$
				& $0.04$ \\
				
				$\phi_{f_1(1510)}$ (rad)
				& $5.30$
				& $5.37$
				& $3.69$
				& $4.28$ \\
				
				$\phi_{f_1(1970)}$ (rad)
				& $1.64$
				& $1.39$
				& $3.62$
				& $4.66$ \\
				
				$\phi_{f_1(2310)}$ (rad)
				& $2.34$
				& $1.76$
				& $5.74$
				& $1.14$ \\
				
			\end{tabular}
		\end{ruledtabular}
	\end{table*}

	\begin{figure}[htbp]
		\centering
		\includegraphics[width=0.3\textwidth]{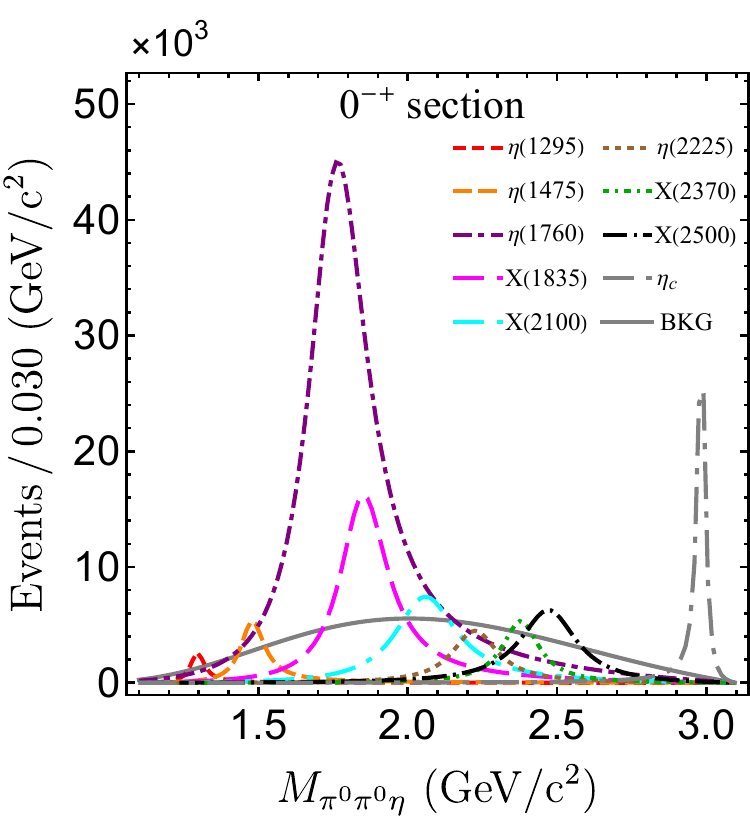}
		\includegraphics[width=0.3\textwidth]{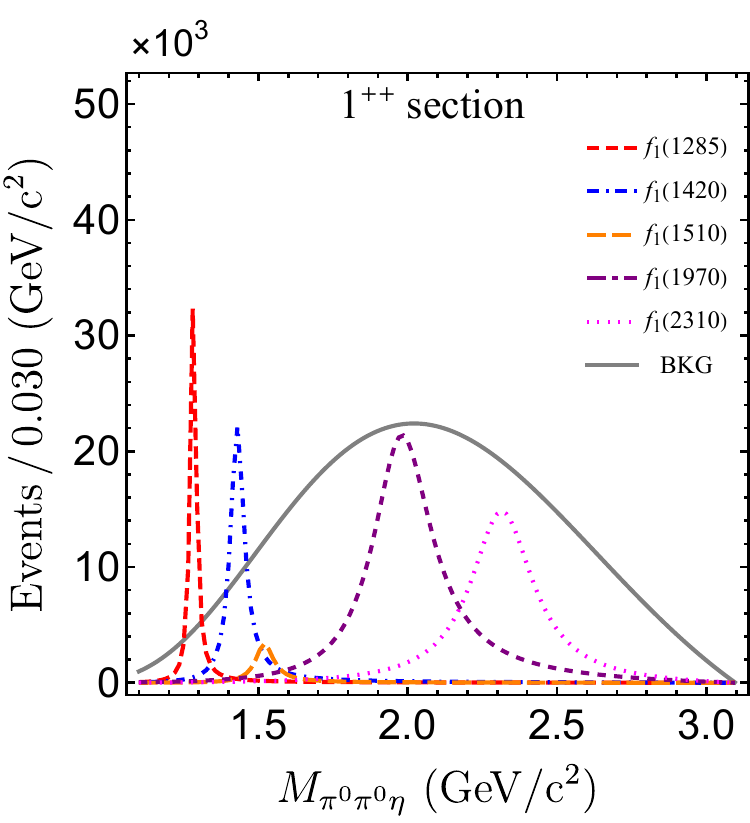}
		\caption{The individual component contributions for $J/\psi\to\gamma\pi^0\pi^0\eta$ in the full mass region: (a) $J^{PC}=0^{-+}$ pseudoscalar contribution; (b) $J^{PC}=1^{++}$ axial-vector contribution.}
		\label{fig:pipieta_section}
	\end{figure}

	\begin{figure}[htbp]
		\centering
		\includegraphics[width=0.3\textwidth]{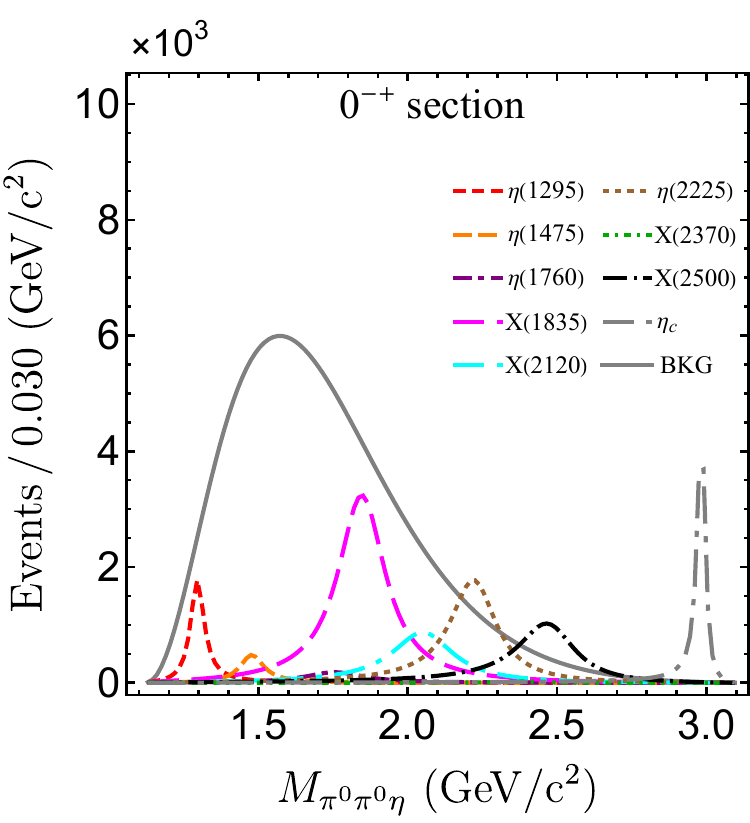}
		\includegraphics[width=0.3\textwidth]{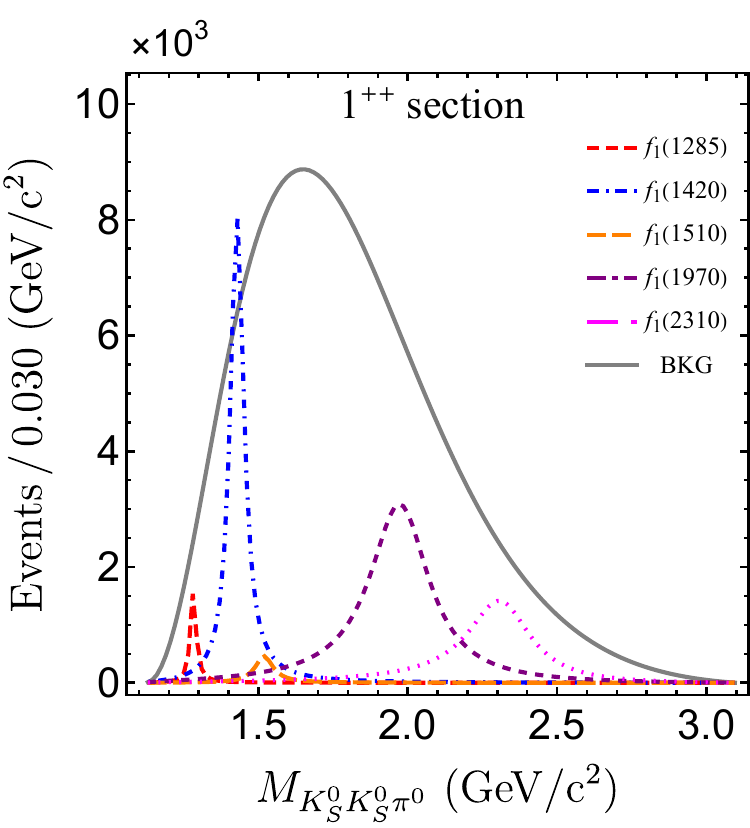}
		\caption{The individual component contributions for
			$J/\psi\to\gamma\pi^0\pi^0\eta$ in the $a_0(980)$ signal region: (a) $J^{PC}=0^{-+}$ pseudoscalar contribution;	(b) $J^{PC}=1^{++}$ axial-vector contribution.}
		\label{fig:pipieta_a0_section}
	\end{figure}

	\begin{figure}[htbp]
		\centering
		\includegraphics[width=0.3\textwidth]{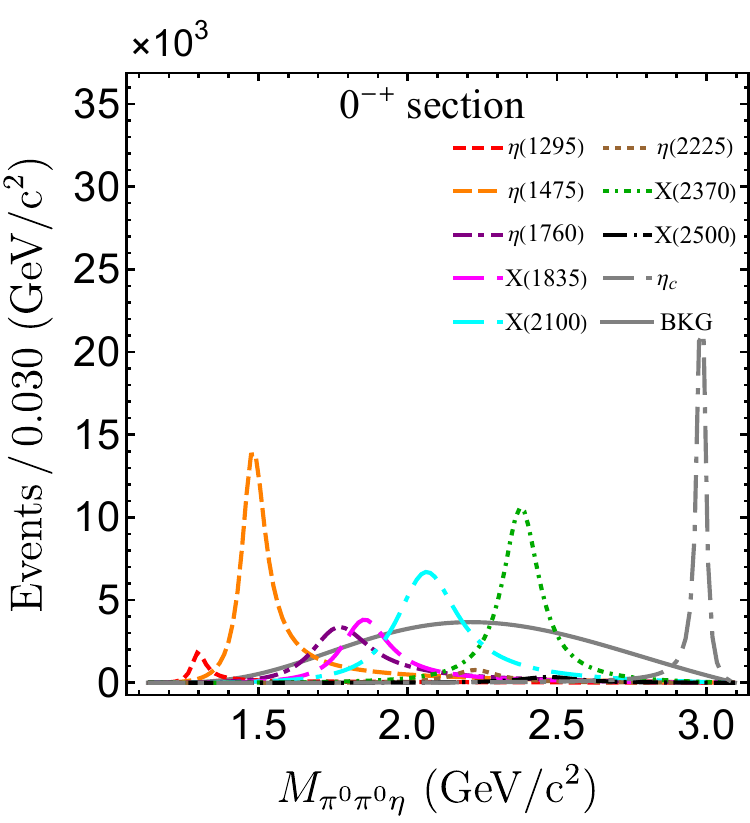}
		\includegraphics[width=0.3\textwidth]{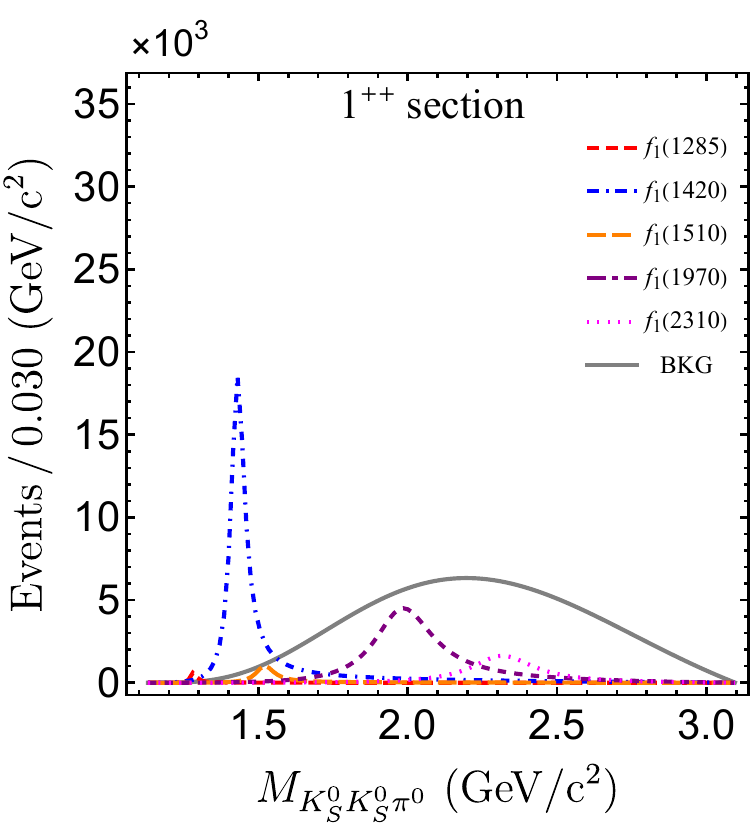}
		\caption{The individual component contributions for $J/\psi\to\gamma K_S^0K_S^0\pi^0$ in the full mass region: (a) $J^{PC}=0^{-+}$ pseudoscalar contribution;	(b) $J^{PC}=1^{++}$ axial-vector contribution.}
		\label{fig:kkpi_section}
	\end{figure}

\end{document}